# Atomic real-space perspective of light-field-driven currents in graphene

Yuya Morimoto[1,2]*, Yasushi Shinohara[3,4] †, Kenichi L. Ishikawa[3-5], and Peter Hommelhoff[1]

[1] *Laser Physics, Department of Physics, Friedrich-Alexander-Universität Erlangen-Nürnberg (FAU), Staudtstraße 1, 91058, Erlangen, Germany*

[2] *Ultrashort Electron Beam Science RIKEN Hakubi research team, RIKEN Cluster for Pioneering Research (CPR), RIKEN Center for Advanced Photonics (RAP), 2-1 Hirosawa, Wako, Saitama 351-0198, Japan*

[3] *Photon Science Center, Graduate School of Engineering, The University of Tokyo, 7-3-1 Hongo, Bunkyo-ku, Tokyo 113-8656, Japan*

[4] *Department of Nuclear Engineering and Management, Graduate School of Engineering, The University of Tokyo, 7-3-1 Hongo, Bunkyo-ku, Tokyo 113-8656, Japan*

[5] *Research Institute for Photon Science and Laser Technology, The University of Tokyo, 7-3-1 Hongo, Bunkyo-ku, Tokyo 113-0033, Japan*

*yuya.morimoto@riken.jp

†Current address: NTT Basic Research Laboratories, 3-1 Morinosato Wakamiya, Atsugi, Kanagawa 243-0198, Japan

Dated: December 9, 2021

**Abstract**

When graphene is exposed to a strong few-cycle optical field, a directional electric current can be induced depending on the carrier-envelope phase of the field. This phenomenon has successfully been explained by the charge dynamics in reciprocal space, namely an asymmetry in the conduction band population left after the laser excitation. However, the corresponding real-space perspective has not been explored so far although it could yield knowledge about the atomic origin of the macroscopic currents. In this work, by adapting the nearest-neighbor tight-binding model including overlap integrals and the semiconductor Bloch equation, we reveal the spatial distributions of the light-field-driven currents on the atomic scale and show how they are related to the light-induced changes of charge densities. The atomic-scale currents flow dominantly through the network of the $\pi$ bonds and are the strongest at the bonds parallel to the field polarization, where an increase of the charge density is observed. The real-space maps of the currents and changes in charge densities are elucidated using simple symmetries connecting real and reciprocal space. We also discuss the strong-field-driven Rabi oscillations appearing in the atomic-scale charge densities. This work highlights the importance of real-space measurements and stimulates future time-resolved atomic-scale experimental studies with high-energy electrons or X-rays, for examples.

1. Introduction

Lightwave electronics, the ultrafast generation and control of macroscopic electrical currents by the waveform of ultrashort laser pulses, holds promise for future high-speed signal processing[1,2]. In experiments, quasi-dc residual currents after the laser excitation were observed. A wide variety of materials has been used as media, such as $SiO_2$[3–5], $CaF_2$[4,6], $Al_2O_3$[4], GaN[5,7,8] and $HfO_2$[5]. Among them, graphene shows its unique character both experimentally[9–13] and theoretically[14,15]. Due to the weak screening effect and the high laser damage threshold, strong light fields can be applied to the single atomic layer of carbon atoms[9–13]. The interaction with strong light fields is well elucidated by considering only two electronic structures, namely $\pi$ and $\pi^*$ bands, allowing to give clear physical interpretations on the origins of the strong-field-driven high-order harmonics and residual electric currents [9–13,15–18], and profound insights and predictions[14,15,18–24], for example, the emergence of the valley polarization induced by two counter-rotating circularly polarized fields [22] and the Berry phase interferometry in reciprocal space [19].

Commonly, the strong-field-driven electron dynamics in graphene is considered in the reciprocal space[9–17,19] [18,20–26]. The electronic excitation usually occurs at around the special points of *K* and *K'* where the quasi-linear dispersion curves of the two bands cross and form the Dirac cones. The time dependent charge dynamics follows the acceleration theorem, that is, $\boldsymbol{k}(t) = \boldsymbol{k}_0 + e\boldsymbol{A}(t)/\hbar$, where $\boldsymbol{k}_0$ is the wavevector of the charge before the interaction with the laser, *e* the elementary charge, $\boldsymbol{A}(t)$ the vector potential of the laser field, and $\hbar$ the reduced Planck constant. The residual electric current induced by a spatially-asymmetric few-cycle laser pulse excitation is attributed to the momentum-asymmetric population of the conduction band ($\pi^*$ state) produced via field-cycle-driven Landau–Zener–Stückelberg interference[9]. However, a discussion on the corresponding real-space pictures is lacking although the currents and the associated charge densities visualized on atomic scale should give deep chemical insights and perspectives at the level of atoms and bonds.

In this work, we theoretically study the light-driven residual currents and changes in charge densities in pristine graphene on the atomic scale. We employ the two-bands ($\pi$ and $\pi^*$ state) nearest-neighbor tight-binding model[9–13,15,18–20,22,23,26–33] . Importantly, by following Ref.[32] we include the overlap integral between neighboring atomic orbitals, which can be omitted in the reciprocal-space discussion [9–13,15] [18–20,22,23,26], otherwise the charge density between atomic sites cannot be described properly. The light-graphene interaction is calculated by the semiconductor Bloch equations. This theoretical treatment containing the essential physics in the simplest possible manner is accurate enough to reproduce experimental observations quantitatively and allows to provide deep insights. We find that the residual currents flow mainly through the hexagonal framework of the $\pi$ bonds rather than the center of hexagon. An increase of the charge density is observed at the bonds where large currents are driven. The symmetry breaking in the reciprocal space induced by the few-cycle excitation is not found in the real-space charge density. Instead, the directional flow of currents breaks the real-space inversion symmetry. We introduce a qualitative discussion based on the reciprocal-space symmetries of graphene, which allows to make a clear connection between the atomic-scale perspectives of this work and the ordinal reciprocal-space pictures. The real-space charge densities contain information about the history of laser excitation such as multi-photon absorption and Rabi oscillations, which underlines the importance of real-space measurements.

## 2. Theoretical model

We consider the $\pi$ and $\pi^*$ bands formed by the $2p_z$ atomic orbitals. Even though the two level approximation neglects the contributions from electrons in 1s and the sp$^2$-hybridization states[25,34], its validity has been demonstrated by the ab-initio calculations even at a high laser field strength of 5 V/nm[25]. Moreover, this simple framework allows us to interpret simulation results concisely and deeply. The electronic eigenstates of graphene $\psi_{\boldsymbol{k}}(\boldsymbol{r})$ are constructed as linear combinations of the tight-binding Bloch functions of the two carbon atoms A1 and A2 in a unit cell, $\Phi_{\boldsymbol{k},j}(\boldsymbol{r})$ ($j = $ A1, A2) which are expressed as

$$\Phi_{\boldsymbol{k},j}(\boldsymbol{r}) = \sum_{\boldsymbol{R}_j}^{N} e^{i\boldsymbol{k}\cdot\boldsymbol{R}_j} \varphi_{2p_z}(\boldsymbol{r} - \boldsymbol{R}_j), \qquad (1)$$

where $\boldsymbol{R}_j$ is the position of the $j$-th atom, $\varphi_{2p_z}$ the wavefunction of a carbon atom in the $2p_z$ state (see below) and $N$ the total number of unit cells. The field-free eigenstates of graphene $\psi_{\boldsymbol{k}}(\boldsymbol{r})$ are obtained by solving the secular equation $\det(H - \varepsilon S) = 0$, where $\varepsilon$ is the eigen energy. In the nearest-neighbor approximation[32], the matrix elements of $H$ and $S$ are written as

$$H = \begin{pmatrix} \varepsilon_{2p} & \gamma f(\boldsymbol{k}) \\ \gamma f(\boldsymbol{k})^* & \varepsilon_{2p} \end{pmatrix}, \quad S = \begin{pmatrix} 1 & sf(\boldsymbol{k}) \\ sf(\boldsymbol{k})^* & 1 \end{pmatrix}, \qquad (2)$$

where $\varepsilon_{2p}$ is the eigen energy of the $2p_z$ orbital, $\gamma = \langle \varphi_{2p_z}(\boldsymbol{r} - \boldsymbol{R}_{A1})|\mathcal{H}|\varphi_{2p_z}(\boldsymbol{r} - \boldsymbol{R}_{A1})\rangle$ the hopping integral where $\mathcal{H}$ is the Hamiltonian of graphene (note that the explicit form of $\mathcal{H}$ is often not discussed [32]), and $s = \langle \varphi_{2p_z}(\boldsymbol{r} - \boldsymbol{R}_{A1})|\varphi_{2p_z}(\boldsymbol{r} - \boldsymbol{R}_{A2})\rangle$ the overlap integral of the neighboring atomic orbitals. With the spatial coordinate shown in Fig. 1(a), where graphene lies in the $xy$ plane, $f(\boldsymbol{k})$ is given by

$$f(\boldsymbol{k}) = \exp\left(\frac{ik_x a}{\sqrt{3}}\right) + 2\exp\left(-\frac{ik_x a}{2\sqrt{3}}\right)\cos\left(\frac{k_y a}{2}\right), \qquad (3)$$

where $a = 2.46$ Å. The solution of the secular equation gives the field-free electronic eigenstates of graphene as

$$\psi_{\boldsymbol{k},\pm}(\boldsymbol{r}) = \frac{1}{\sqrt{2(1 \pm s|f(\boldsymbol{k})|)}}\{\Phi_{\boldsymbol{k},A1}(\boldsymbol{r}) \pm e^{-i\phi(\boldsymbol{k})}\Phi_{\boldsymbol{k},A2}(\boldsymbol{r})\}. \qquad (4)$$

The wavefunction with the plus sign corresponds to the $\pi$ state and the minus sign stands for the $\pi^*$ state. These eigenstates are normalized to $\int |\psi_{\boldsymbol{k},\pm}(\boldsymbol{r})|^2 d\boldsymbol{r} = 1$, where the spatial integral is taken over the unit cell of graphene. $\phi(\boldsymbol{k})$ in Eq. (4) is defined as

$$e^{i\phi(\boldsymbol{k})} = \frac{f(\boldsymbol{k})}{|f(\boldsymbol{k})|}. \qquad (5)$$

The eigen energies for the two states are given by

$$\varepsilon_{\pm}(\mathbf{k}) = \frac{\varepsilon_{2p} \pm \gamma |f(\mathbf{k})|}{1 \pm s|f(\mathbf{k})|}. \tag{6}$$

Following [32], we use the parameters of $\varepsilon_{2p} = 0$, $\gamma = -3.033$ eV, and $s = 0.129$ to reproduce the energy bands given by the first principles calculation [32]. The band gap $\Delta\varepsilon(\mathbf{k}) = \varepsilon_{-}(\mathbf{k}) - \varepsilon_{+}(\mathbf{k})$ is shown in Fig. 1(b). At the $K$ and $K'$ points, the band gap is zero because of $f(\mathbf{k}) = 0$. Small energy gaps (blue region) can be found along the five lines connecting the symmetric points, namely $K$-$M$-$K'$, $K$-$M'$, $K$-$M''$, $K'$-$M'$ and $K'$-$M''$. We note that the overlap integral of the neighboring atomic orbitals $s = \langle \varphi_{2p_z}(\mathbf{r} - \mathbf{R}_{A1}) | \varphi_{2p_z}(\mathbf{r} - \mathbf{R}_{A2}) \rangle$ is often neglected and in fact not required in the reciprocal-space discussion [9–13,15] [18–20,22,23,26] but vital for the real-space discussion using the atomic orbitals $\varphi_{2p_z}(\mathbf{r})$. By following Ref.[35], we describe the atomic orbital $\varphi_{2p_z}(\mathbf{r})$ with a Slater-type orbital with an effective nuclear charge $Z$,

$$\varphi_{2p_z}(\mathbf{r}) = Zr \cos\theta \exp\left(-\frac{Z}{2}r\right), \tag{7}$$

where the origin of $\mathbf{r}$ is taken at the atomic nucleus and $\theta$ is the polar angle of $\mathbf{r}$ measured from the $z$-axis. Although the value of the overlap integral becomes 0.129 at $Z = 4.02$, we find significant changes of the total charge following the optical excitation. Therefore, we instead choose $Z = 3.13$ (the value of the overlap integral is 0.258) such that the total charge densities of the $\pi$ and $\pi^*$ states integrated over both reciprocal-space and $\mathbf{r}$-space are identical, that is, $\int \int_{BZ} |\psi_{\mathbf{k}_0,+}(\mathbf{r})|^2 d\mathbf{r} d\mathbf{k}_0 = \int \int_{BZ} |\psi_{\mathbf{k}_0,-}(\mathbf{r})|^2 d\mathbf{r} d\mathbf{k}_0$, where BZ stands for the first Brillouin zone. Left and right panels of Fig. 1(a) show two-dimensional charge densities $\rho_{\pm}(x,y) = \int \int_{BZ} |\psi_{\mathbf{k}_0\pm}(\mathbf{r})|^2 dz d\mathbf{k}_0$ for the $\pi$ and $\pi^*$ states, respectively, where $z$ is the real-space coordinate perpendicular to the plane. We employ 128×128 grids for discretizing the first Brillouin zone. In the bonding $\pi$ state (left panel), the charge density is delocalized while the charge density in the anti-bonding $\pi^*$ state (right panel) is localized near the atomic sites. Therefore, the electronic excitation to the $\pi^*$ state normally increases the charge densities at the nuclei and reduces those of the bonds.

The interaction with a laser field is calculated with the semiconductor Bloch equations using the Houston basis within the velocity gauge[26,36–39]. The semiconductor Bloch equation used in this study is derived from the time-dependent Schrödinger equation with the dipole approximation [36],

$$i\hbar \frac{\partial}{\partial t} \Psi_{\bm{k}_0}(\bm{r},t) = h(t)\Psi_{\bm{k}_0}(\bm{r},t), \tag{8}$$

where $h(t) = h_0 - i\hbar e \bm{A}(t)\cdot \bm{\nabla} + e^2 A^2(t)/2$ and $h_0$ is the Hamiltonian without external fields, $h_0 \psi_{\bm{k},\pm}(\bm{r}) = \epsilon_\pm(\bm{k})\psi_{\bm{k},\pm}(\bm{r})$. We describe the time-dependent wavefunction for an electron with the initial lattice momentum $\bm{k}_0$ as

$$\Psi_{\bm{k}_0}(\bm{r},t) = \sum_{n=+,-} c_n(\bm{k}_0,t) \exp\left(-\frac{ie\bm{A}(t)\cdot \bm{r}}{\hbar}\right) \psi_{\bm{k}_0+e\bm{A}(t)/\hbar,n}(\bm{r}), \tag{9}$$

where the coefficients $c_n$ for $n = +,-$ are determined by solving the Schrödinger equation. By introducing the density matrix $\rho_{mn}(\bm{k}_0,t) = c_m^*(\bm{k}_0,t)c_n(\bm{k}_0,t)$, a differential equation for $\rho_{mn}(\bm{k}_0,t)$ is obtained from Eq. (8) as

$$i\hbar \frac{\partial \rho_{mn}(\bm{k}_0,t)}{\partial t} = \left(\varepsilon_n(\bm{k}_0 + e\bm{A}(t)/\hbar) - \varepsilon_m(\bm{k}_0 + \bm{A}e(t)/\hbar)\right)\rho_{mn}(\bm{k}_0,t)$$

$$-\bm{E}(t)\cdot [\bm{d}_{mn}(\bm{k}_0 + e\bm{A}(t)/\hbar)\rho_{mm}(\bm{k}_0,t) - \bm{d}_{mn}^*(\bm{k}_0 + e\bm{A}(t)/\hbar)\rho_{nn}(\bm{k}_0,t)], \tag{10}$$

where $\bm{E}(t)$ is the laser electric field. Equation (10) is a form of the semiconductor Bloch equation [26,36–39]. Relaxation and dephasing effects are neglected as in Refs.[9,15]. This equation is solved numerically by the fourth order Runge-Kutta method with 2.4 as (0.1 au) time steps or less when necessary, to obtain converged results. $\bm{d}_{+-}(\bm{k})$ in Eq. (10) is the dipole matrix element whose *x* and *y* components are given by

$$d_{+-,x}(\bm{k}) = \frac{1}{2\sqrt{3}} \frac{a}{|f(\bm{k})|^2} \sqrt{\frac{1-s|f(\bm{k})|}{1+s|f(\bm{k})|}} \left\{\cos(ak_y) - \cos\left(\frac{\sqrt{3}ak_x}{2}\right)\cos\left(\frac{ak_y}{2}\right)\right\},$$

$$d_{+-,y}(\bm{k}) = \frac{1}{2} \frac{a}{|f(\bm{k})|^2} \sqrt{\frac{1-s|f(\bm{k})|}{1+s|f(\bm{k})|}} \sin\left(\frac{\sqrt{3}ak_x}{2}\right)\sin\left(\frac{ak_y}{2}\right), \tag{11}$$

respectively. A difference from the dipole matrix elements for the case of $s=0$ of Ref[26] is the appearance of the factor $\sqrt{(1-s|f(\bm{k})|)/(1+s|f(\bm{k})|)}$. However, because carrier excitations dominantly occur at $\bm{k}$ near the K and K' points, where $f(\bm{k}) \simeq 0$, the inclusion of the overlap integral causes little influence on the reciprocal-space dynamics. The magnitudes $|d_{+-,x}(\bm{k})|$ and $|d_{+-,y}(\bm{k})|$ are plotted in Figs. 1(c). While $|d_{+-,x}(\bm{k})|$ has large values along the K-M-K' line, $|d_{+-,y}(\bm{k})|$ is large along the 4 lines of K-M' (M'') and *K'- M'* (*M''*). This ***k***-dependence for the two orthogonal polarization directions dictates the sign and the spatial distribution of the current densities as well as the laser-driven changes in charge densities.

Experimentally, the light-driven electric current is measured at a time scale much longer than the laser pulses, typically on the order of nanoseconds [9]. Therefore, we focus on the quasi-stationary state and the residual current after the laser excitation. To this end, we solve the semiconductor Bloch equations [Eq. (10)] with a given waveform of $E(t)$ (see below) and obtain the density matrix after the laser excitation, $\rho_{mn}(\mathbf{k}_0, t = \tau)$, where $\tau$ is the time when the laser field intensity becomes sufficiently weak [see Eq. (14) below]. We then take the diagonal terms of the density matrix after the laser excitation, $\rho_{++}$ and $\rho_{--}$, both of which contribute to the time-constant (i.e. residual) current and charge densities. The off-diagonal elements $\rho_{+-}$ and $\rho_{-+}$, on the other hand, induce time-oscillatory dynamics at the time scales similar to the optical cycle and their time-averaged contribution is zero. Therefore, the contributions from the off-diagonal elements are neglected in this work. For convenience, we describe the reciprocal-space populations in the $\pi$ and $\pi^*$ states after the laser excitation as $n_\pi^{\text{After}}(\mathbf{k}_0) = \rho_{++}(\mathbf{k}_0, t = \tau)$ and $n_{\pi^*}^{\text{After}}(\mathbf{k}_0) = \rho_{--}(\mathbf{k}_0, t = \tau)$. Our theoretical framework omits the electron-electron and the electron-phonon scatterings, thus the total charges is conserved, $n_\pi^{\text{After}}(\mathbf{k}_0) + n_{\pi^*}^{\text{After}}(\mathbf{k}_0) = 1$ for each $\mathbf{k}_0$. By using these expressions, the residual charge current densities along the *x* and *y* directions $J_x^{\text{DC}}(\mathbf{r})$ and $J_y^{\text{DC}}(\mathbf{r})$ and the residual charge density $\rho^{\text{DC}}(\mathbf{r})$ are given by

$$J_x^{\text{DC}}(\mathbf{r}) \propto \int d\mathbf{k}_0 \left[ (1 - n_{\pi^*}^{\text{After}}(\mathbf{k}_0)) \text{Im} \left\{ \psi_{\mathbf{k}_0,+}^*(\mathbf{r}) \frac{\partial \psi_{\mathbf{k}_0,+}(\mathbf{r})}{\partial x} \right\} + n_{\pi^*}^{\text{After}}(\mathbf{k}_0) \text{Im} \left\{ \psi_{\mathbf{k}_0,-}^*(\mathbf{r}) \frac{\partial \psi_{\mathbf{k}_0,-}(\mathbf{r})}{\partial x} \right\} \right],$$

$$J_y^{\text{DC}}(\mathbf{r}) \propto \int d\mathbf{k}_0 \left[ (1 - n_{\pi^*}^{\text{After}}(\mathbf{k}_0)) \text{Im} \left\{ \psi_{\mathbf{k}_0,+}^*(\mathbf{r}) \frac{\partial \psi_{\mathbf{k}_0,+}(\mathbf{r})}{\partial y} \right\} + n_{\pi^*}^{\text{After}}(\mathbf{k}_0) \text{Im} \left\{ \psi_{\mathbf{k}_0,-}^*(\mathbf{r}) \frac{\partial \psi_{\mathbf{k}_0,-}(\mathbf{r})}{\partial y} \right\} \right], \quad (12)$$

and

$$\rho^{\text{DC}}(\mathbf{r}) = \frac{2}{S_{\text{BZ}}} \int d\mathbf{k}_0 \left\{ \left(1 - n_{\pi^*}^{\text{After}}(\mathbf{k}_0)\right) |\psi_{\mathbf{k}_0,+}(\mathbf{r})|^2 + n_{\pi^*}^{\text{After}}(\mathbf{k}_0) |\psi_{\mathbf{k}_0,-}(\mathbf{r})|^2 \right\},$$

$$S_{\text{BZ}} = \frac{(2\pi)^2}{\sqrt{3}a^2/2}, \quad (13)$$

respectively. When integrated over the space, $J_x^{\text{DC}}(\mathbf{r})$ and $J_y^{\text{DC}}(\mathbf{r})$ give the residual intraband currents along *x* and *y* directions observed in experiments[9–13]. In the following, we discuss two-dimensional current densities which are given by integrals over the *z*-axis, $J_x^{\text{DC}}(x,y) = \int J_x^{\text{DC}}(\mathbf{r}) dz$ and $J_y^{\text{DC}}(x,y) = \int J_y^{\text{DC}}(\mathbf{r}) dz$. Because the changes in charge densities are much smaller compared to the initial charge density

(of the order of 0.1% with an excitation field of V/nm amplitude), we discuss the changes defined by

$$\Delta\rho^{DC}(x,y) = \int\left\{\rho^{DC}(\boldsymbol{r}) - \int|\psi_{\boldsymbol{k}_0,+}(\boldsymbol{r})|^2 d\boldsymbol{k}_0\right\}dz = \iint n_{\pi^*}^{\text{After}}(\boldsymbol{k}_0)\left\{-|\psi_{\boldsymbol{k}_0,+}(\boldsymbol{r})|^2 + |\psi_{\boldsymbol{k}_0,-}(\boldsymbol{r})|^2\right\}d\boldsymbol{k}_0 dz.$$

Before presenting numerical results, we briefly discuss and confirm the validity of our theoretical model. The equation of continuity for the residual charge and current densities reads $\boldsymbol{\nabla}\cdot\boldsymbol{J}^{DC}(\boldsymbol{r}) = \frac{\partial J_x^{DC}(\boldsymbol{r})}{\partial x} + \frac{\partial J_y^{DC}(\boldsymbol{r})}{\partial y} = \partial\rho^{DC}(\boldsymbol{r})/\partial t = 0$ at each $\boldsymbol{r}$. This leads to the integral form of $\iint_S \boldsymbol{J}^{DC}(\boldsymbol{r})\cdot d\boldsymbol{S} = 0$ for any closed surface $S$. For example, when $S$ is the surfaces of a rectangular parallelepiped with eight edges of (0, 0, $\pm L_z/2$), ($L_x$, 0, $\pm L_z/2$), (0, $a$, $\pm L_z/2$) and ($L_x$, $a$, $\pm L_z/2$) with $L_z\to+\infty$, the integral form becomes $\int_0^a J_x^{DC}(x=0,y)dy - \int_0^a J_x^{DC}(x=L_x,y)dy = 0$ for any $L_x$ due to the lattice periodicity. Therefore, the equation of continuity requests that the value of $\int_0^a J_x^{DC}(x,y)dy$ is identical at any $x$, which is indeed confirmed in our results at any field strength within ~10% error. The small error attributes to the use of the nearest-neighbor tight-binding approximation together with the atomic wavefunction $\varphi_{2p_z}(\boldsymbol{r})$ [Eq. (7)] having very small but non-zero overlap with higher-order neighbors.

## 3. Results and discussion

We consider the excitation with a laser electric field of the form of

$$E(t) = F_0 \sin^4\left(\frac{\pi t}{\tau}\right)\cos(\omega t + \varphi_{\text{CEP}}), \tag{14}$$

where $F_0$ is the peak field amplitude, $\tau$ gives pulse duration, $\omega$ is the angular frequency of the field and $\varphi_{\text{CEP}}$ is the carrier-envelope phase by which the magnitudes and the direction of the residual current are controlled. The sin$^4$ envelope is used rather than the sin$^2$ envelope, in order to avoid the high frequency tail of the excitation field spectrum[40]. We choose $\omega = 2\pi\times0.375$ PHz corresponding to the wavelength of 800 nm. The pulse duration is adjusted to be $\tau = 21$ fs, which corresponds to the full-width-at-half-maximum (FWHM) duration of 5.5 fs, in order to obtain the best match with the experimental results[9]. An example of the waveform with $\varphi_{\text{CEP}} = \pi/2$ is plotted in Fig. 1(d) as a red curve. We consider the peak field amplitude $F_0$ up to 4 V/nm which is below the reported damage threshold (4.5 V/nm) [41]. Figure 1(e) shows the total residual currents, i.e., the spatial integrals of $J_x^{DC}(\boldsymbol{r})$ and $J_y^{DC}(\boldsymbol{r})$, driven by the linearly polarized fields along the $x$ and $y$ axes. They match the experimental results[9] (open circles) almost perfectly, further demonstrating the validity of our theoretical model including the non-zero overlap integral.

Below, for convenience, we denote the fields polarized along *x* and *y* axes as $E_x$ and $E_y$, respectively. The non-zero residual currents are observed only along the polarization direction. The direction of the residual current at $\varphi_{CEP} = -\pi/2$ (blue curve) is reversed as compared to that of $\varphi_{CEP} = \pi/2$ (red curve). No residual currents are found at $\varphi_{CEP} = 0$ (green curve). Excitation fields with the two polarization directions ($E_x$ and $E_y$) give nearly the same amount of currents, also agreeing with the experiment[9].

Figure 2 summarizes the results of the $E_x$ excitation with $\varphi_{CEP} = \pi/2$. The panels in Fig. 2(a) show the reciprocal-space population of the $\pi^*$ state, $n_{\pi^*}^{After}(k_x, k_y)$, around the *K* point, calculated at variable peak field amplitudes ($F_0$) indicated above each panel. The populations around the *K'* point (not shown) are given by flipping over these plots with respect to $k_y = 0$. Very complex structures with interferences and multiple nodes are observed. The ring patterns (see orange curves in the panel for 2 V/nm) suggest that the excitation dominantly occurs through the single and multi-photon processes. Small left-right asymmetries with respect to $k_x = 14.7$ nm$^{-1}$, i.e., the $k_x$ value of the *K* point, are clearly visible at high field strengths. It is the origin of the macroscopic residual current[9–13]. The plots in Fig. 2(a) are very close to those in Ref[9], where the overlap integral is neglected, confirming the validity of our simulations.

Figure 2(b) shows the laser-induced change of the real-space charge densities $\Delta\rho^{DC}(x, y)$, which are much simpler as compared to the reciprocal-space densities in Fig. 2(a). We observe 4 main features. First, the change of densities at the nuclei (black circles filled with yellow) are always positive (red). Second, the density changes at the bonds tilted by ±60 degrees from the *x* axis are always negative (blue). These two features can be attributed to the excitation to the anti-bonding $\pi^*$ state, see also Fig. 1(a). Third, the density change at the bonds parallel to the *x* axis shows a completely different behavior and a nontrivial dependence on the field strength. It is positive (red) at 0.5 V/nm, negative (blue) at 1.0 V/nm and positive again at higher field strengths. As disused below, this behavior can be explained by the interplay of a $k_0$-selective excitation and Rabi oscillations. Fourth, the charge densities do not show any left-right asymmetry. For example, the charge densities of all the atomic sites for a given $F_0$ are identical. Thus, the symmetry breaking appearing in the reciprocal space is not seen in the real-space charge densities. Below, we discuss its origin based on symmetries. Instead, the inversion symmetry in the real space is broken by the directional current, as we show now.

Figure 2(c) shows the real-space current densities flowing along the *x*-axis, $J_x^{DC}(x, y)$. The current densities have nearly the same pattern for all field amplitudes except the differences in signs and magnitudes. The magnitudes of the current densities are the largest at the bonds parallel to the electric field (along *x*- axis), where the charge density is enhanced by the laser excitation, see above and Fig. 2(b). Therefore, the increase of the charge density can be regarded as a formation of a channel for the current flow.

The macroscopic currents perpendicular to the field polarization were not observed in experiments[9–13], showing that the spatial integral of $J_y^{DC}(x,y)$ is zero, $\iint J_y^{DC}(x,y)dxdy = 0$. This is a natural consequence that the symmetry along the *y*-axis is not broken by the $E_x$ excitation, and is indeed confirmed in all our results (not shown). However, the spatially-resolved current density itself $J_y^{DC}(x,y)$ does not have to be zero. The vectorial representation of the current is given in the left panel of Fig. 1(f) for $F_0 = 3$ V/nm. We find significant contributions of $J_y^{DC}(x,y)$ (vertical component) at around $x = -0.4$ and 1.8 Å. The current flows along the hexagonal atomic framework rather than through the center of the hexagon. The current passes through the bond parallel to the field and is split into the two pathways along the bonds oriented at 60 degrees.

In order to discuss these findings in more details, we focus on six special points in real space, namely, the two atomic nuclei (A1, A2), the three bonds (B1, B2 and B3) and the center of the hexagon (C), see right panel of Fig. 1(a). The upper panels of Fig. 3(a) show the changes in charge densities $\Delta\rho^{DC}(x,y)$ (see above for the definition) at the six spatial points, which are normalized by the density before laser excitation. At each spatial point, the change of charge density induced by the field of $\varphi_{CEP} = \pi/2$ (red curves) is identical to that by the field of $\varphi_{CEP} = -\pi/2$ (blue broken curves) within our numerical accuracy (8 digits). This shows that no inversion symmetry breaking occurs in the real-space charge density. Due to this symmetry, A1 and A2 show the identical profiles. B2 and B3 also show the identical results, which is why these pairs are plotted in the same panels. With increasing the peak field amplitudes (horizontal axis), we observe an almost monotonic increase of density at A1, A2 and decrease at B2, B3. In contrast, the charge densities at B1 and C show a completely different dependence on the field amplitude. In these panels, we observe oscillatory features, the clearest at B1.

We explain the observed charge-density changes based on the symmetries connecting real and reciprocal space. Figure 4(a) shows the contributions to the real-space charge-density changes from each momentum $\boldsymbol{k}_0$, that is $\Delta\rho^{DC}(x,y,\boldsymbol{k}_0) = \int \left\{ -\left|\psi_{\boldsymbol{k}_0,+}(\boldsymbol{r})\right|^2 + \left|\psi_{\boldsymbol{k}_0,-}(\boldsymbol{r})\right|^2 \right\} dz$, for the six spatial points. This definition comes from the fact that our theory does not include transitions between different $\boldsymbol{k}_0$ such as electron-electron and electron-phonon scatterings, and accordingly the total amount of charge is conserved at each $\boldsymbol{k}_0$. This quantity represents whether and how much the carrier excitation occurring at $\boldsymbol{k}_0$ increases or decreases the charge density at $(x,y)$. For the atomic sites A1 and A2 (left most panel), the excitation at any $\boldsymbol{k}_0$ increases the charge density. On the other hand, the charge densities of B1, B2, B3 and C strongly depend on $k_0$. In other words, the momentum distributions of the excited carriers in the Brillouin zone $n_{\pi*}^{\text{After}}(k_x, k_y)$ have a strong influence on the charge densities at these points. For example, the excitation at $k_0$ along the K-M-K' line, where $k_x$ is large but $k_y$ is small, increases the charge densities at B1 and C but decreases them at B2 and B3. Conversely, the excitation at $k_0$ along the K-M' line increases the charge densities at B3 and C but decreases them at B1 and B2. We note that this symmetry discussion with Fig. 4(a) is universal for any excitation field strength, wavelength and polarization.

We now return to the case of the $E_x$ excitation and the dipole matrix element of $d_{+-,x}(k)$ in Fig. 1(c). Large values of $|d_{+-,x}(k)|$ can be found at around the K-M-K' line, suggesting that the electronic excitation preferably occurs at around the line. Indeed, the reciprocal-space population of the π* state $n_{\pi*}(k_x, k_y)$ shown in Fig. 2(a) has larger population at $0 < k_y < 8.5$ nm$^{-1}$ as compared to that at $k_y > 8.5$ nm$^{-1}$, where $k_y = 8.5$ nm$^{-1}$ is the $k_y$ value of the K point. The symmetries involved account for the overall increase at B1, C and decrease at B2, B3 seen in Fig. 3(a). The rapid decrease of the density at C occurring at high field amplitude is discussed below. The symmetries in Fig. 4(a) also explain why the inversion symmetry is kept in the real-space charge densities. Although the excitation with the $E_x$ field of $\varphi_{\text{CEP}} = \pi/2$ produces an asymmetric $n_{\pi*}^{\text{After}}(k_x, k_y)$ at around $k_x = 14.7$ nm$^{-1}$, Fig. 4(a) suggests that this momentum-asymmetric population is not translated into any asymmetry in the real-space charge densities. The plots for A1, A2, B1 and C are symmetric at around $k_x = 14.7$ nm$^{-1}$. Even though the plots for B2 and B3 are not symmetric at around $k_x = 14.7$ nm$^{-1}$, the asymmetry around the K point is cancelled by that around the K' point. Therefore, in order to break the inversion symmetry in the real-space charge densities, one may need different excitation probabilities around the K and K' points, i.e., a valley-selective excitation[22].

We can learn more than the reciprocal-space symmetries. The lower panels of Fig. 3(a) show the decompositions of the results in the upper panels with respect to the band gap $\Delta\varepsilon(k)$ at each momentum $k_0$. Here the band gap is counted in units of the photon energy (1.55 eV, corresponding to the 800-nm wavelength). This choice is motivated by the photon-order ring patterns seen in $n_{\pi*}^{\text{After}}(k_x, k_y)$ of Fig. 2(a). The N-photon ($N = 1,2,3,..$) channel here represents the change of charge density originating from $k_0$ satisfying $(N - 0.5)\hbar\omega \leq \Delta\varepsilon(k_0) < (N + 0.5)\hbar\omega$. At a field amplitude above 1 V/nm, the channels of $N \geq 2$ dominate. Therefore, the changes in the charge densities are highly nonlinear. Interestingly, the 1-photon channels (red curves) and 2-photon channels (orange broken lines) show clear oscillations in all the panels. We attribute them to the Rabi oscillations. In the case of continuous wave excitation, the Rabi frequency is given by[42] $\Omega(k_0) = |d_{+-,x}(k_0)|F_0/\hbar$ for each $k_0$. For example, $|d_{+-,x}|$ is 8.5 au at $(k_x, k_y) = $ (14.7 nm$^{-1}$, 7.1 nm$^{-1}$) where the band gap is equal to the $1\hbar\omega$ on the K-M-K' line. The field amplitude giving a $2\pi$ phase for the duration of 5.5 fs (FWHM duration of the laser intensity envelope) is approximately given by $F_0 = 1.7$ V/nm, which agrees well with the oscillation cycles appearing in Fig. 3(a). The magnitudes of the dipole matrix elements for the 2-photon energy gaps are lower [see Fig. 1(b)]. For example, $|d_{+-,x}|$ is 4.6 au at $(k_x, k_y) = $ (14.7 nm$^{-1}$, 5.7 nm$^{-1}$), which is roughly half of that for $N = 1$ and accordingly the field amplitude needed for $2\pi$ phase is twice higher. The twice longer period is indeed seen in the results of lower panels of Fig. 3(a).

The photon-order decomposition also helps to understand the charge density at point C. The sudden decrease of charge density at C occurring at high (>3 V/nm) field strength seen in Fig. 3(a) mainly originates from the $N = 3$ (green curve) and 4 channels (purple curve). In the right most panel of Fig. 4(a), the region of momenta giving a positive density change at C is very narrow at around the lines of K-M-K', K-M' and so on. With increasing field strength, the reciprocal-space populations at larger gaps increase and become broader [see Fig. 2(a)]. Accordingly, the charge density at C decreases at high field strengths.

We now consider current densities at the six special points in real space. The current densities $J_x^{DC}(x, y)$ and their photon-order decompositions are plotted in Fig. 3(b). The current densities at the five spatial points except C show identical dependences on the peak field amplitude, hinting at a common origin of the currents. In order to elucidate this observation, we plot in Fig. 4(b) the contributions to the current density from each momentum $\boldsymbol{k}_0$, $J_x^{DC}(x, y, \boldsymbol{k}_0) = -\int \text{Im}\{\psi_{\boldsymbol{k}_0,+}^*(\boldsymbol{r})\partial\psi_{\boldsymbol{k}_0,+}(\boldsymbol{r})/\partial x\} dz + \int \text{Im}\{\psi_{\boldsymbol{k}_0,-}^*(\boldsymbol{r})\partial\psi_{\boldsymbol{k}_0,-}(\boldsymbol{r})/\partial x\} dz$. Indeed, these reciprocal-space symmetries leading to the currents at A1, A2, and B1 are identical. As has been discussed in Refs. [9,10,12], the left-right asymmetry at around $k_x = 14.7$ nm$^{-1}$ is the key for $J_x^{DC}$. The symmetries for B2 and B3 are different but the average of $k_y > 0$ and $k_y < 0$ gives the same pattern as those for A1, A2 and B1. Therefore, the reciprocal-space symmetry associated with the current densities at the five points (A1, A2, B1-B3) is the same.

On the other hand, the reciprocal-space symmetry for the current at C is different from the others. Accordingly, the dependence on the field amplitude is also unique, contributing the unique behavior in the right most panel of Fig. 3(b). At the peak amplitude in the range of 3.3-3.8 V/nm [see Fig. 1(e)], the direction of the current at C is opposite to those at the other five points. The corresponding real-space vectorial image at 3.8 V/nm, where the total residual current is nearly zero, is shown in the middle panel of Fig. 1(f). A circulation of current through the center of the hexagon occurs. This interesting phenomenon of the ring-like current made inside the atomic hexagon was only found thanks to the real-space analysis of this work.

Considering the oscillatory features of the residual currents with respect to the peak field amplitude [Fig. 3(b)], there might be other field amplitudes above 4 V/nm giving the ring-like currents. On the other hand, the ring-current formation might depend on the theoretical framework and therefore should be confirmed by higher-level theories or ideally by experiments.

The photon-order decompositions of the current densities are plotted in the lower panels of Fig. 3(b). The dominating photon orders are 1 and 2. In Fig. 2(a), the asymmetric population in reciprocal space can indeed be seen mainly inside of the 3rd ring. The dependence on the peak field amplitude is different from that of the charge densities in Fig. 3(a). No signature of Rabi oscillation is seen in Fig. 3(b). This shows that the total density population [Fig. 3(a)] and the asymmetry [i.e., the residual current, Fig. 3(b)] depend on the field amplitude in different ways, suggesting the importance of the real-space measurement of both the probability and current densities.

Finally, we apply the same discussion to the results of the $E_y$ excitation. Figure 5(a) shows the changes in real-space charge density $\Delta\rho^{DC}(x,y)$ at eight peak field amplitudes. In contrast to the case of the $E_x$ excitation shown in Fig. 2(b), the charge density at B1 is always strongly reduced. On the other hand, the decrease of the charge densities at B2 and B3 are much weaker, or they even increase at some field amplitudes. The real-space current densities $J_y^{DC}(x,y)$ are plotted in Fig. 5(c) for the peak field amplitudes of 2 V/nm and 3 V/nm. The atomic-scale charge current is mainly driven through B2 and B3, i.e., the bonds aligned rather parallel to the field polarization direction. Therefore, the charge densities on the passage of the currents are enhanced also with the $E_y$ excitation. We note that the butterfly-like patterns of $J_y^{DC}(x,y)$ can be found in a simulation result of a previous work including many electronic bands[34], suggesting the validity of two-level approximation used here. The magnitude of the dipole matrix element for $E_y$, $|d_{+-,y}(\boldsymbol{k})|$, plotted in the right panel of Fig. 1(c), has large values at around the lines of K-M', K-M'', K'-M' and K'-M''. Accordingly, as shown in Fig. 5(b), the excitation occurs preferably at $k_y > 8.5$ nm$^{-1}$ and $k_y < -8.5$ nm$^{-1}$ in the first Brillouin zone; compare to the case of the $E_x$ excitation in Fig. 2(a), where the opposite holds. Combined with the reciprocal-space symmetries in Fig. 4(a), which are independent of the excitation field polarization and strength, carrier excitations at the momenta around these lines decrease the charge density at B1. The densities at B2 and B3 are kept rather constant because there are both positive and negative contributions at momenta along these lines. Figure 5(d) shows the changes in charge densities with respect to the peak field amplitudes. Oscillatory features originating from Rabi oscillations are clearly seen. The current densities at the atomic sites A1, A2 and the bonds B1-B3 show the same dependence on the peak field amplitude, see Fig. 5(e).

## 4. Conclusion

In conclusion, we have investigated the atomic-scale light-field-driven residual currents in pristine graphene. The currents flow dominantly through the network of the π bonds. Even though the electronic excitation to the anti-bonding π* state generally decreases the charge densities at bonds, the increase of the densities occurs at the bonds where large currents are driven, which can be interpreted as a channel formation for currents. There are special peak field amplitudes that drive currents at the center of hexagon in the opposite direction as to those at the nuclei and bonds in a way that the ring-like current is formed. The residual currents and the laser-driven changes in the charge densities were elucidated based on the simple reciprocal-space symmetries. We showed that strong-field-driven Rabi oscillations can be investigated through the measurement of the real-space charge densities.

The real-space quantities discussed in this work should be observable with existing experimental approaches; atomic-scale charge densities can be precisely determined by X-ray or electron diffraction[43,44] and atomic-resolution electron microscopy or X-ray scattering can capture the real-space current densities, as shown by recent theoretical studies[45,46]. No extreme temporal resolution such as attoseconds[47–50] is required since the time scale of the carrier relaxation is longer than 10 fs[51–55]. Our work thus guides future experimental studies on the microscopic dynamics occurring in the media of the light-wave electronics beyond the macroscopic measurements.

**Acknowledgement**

The experimental data[9] plotted in Fig. 1(e) was provided by courtesy of Tobias Boolakee and Takuya Higuchi. This work was supported by the Gordon and Betty Moore Foundation (GBMF) through Grant No. GBMF4744, "Accelerator on a Chip International Program-ACHIP," the European Research Council (ERC) Advanced Grant No. 884217 "AccelOnChip", the FAU Emerging Talents Initiative, Ministry of Education, Culture, Sports, Science and Technology (MEXT Q-LEAP JPMXS0118067246); Japan Society for the Promotion of Science (18K14145, 19H00869, 19H02623, JP20H05670); Japan Science and Technology Agency(Center of Innovation Program, CREST JPMJCR15N1, Research and Education Consortium).

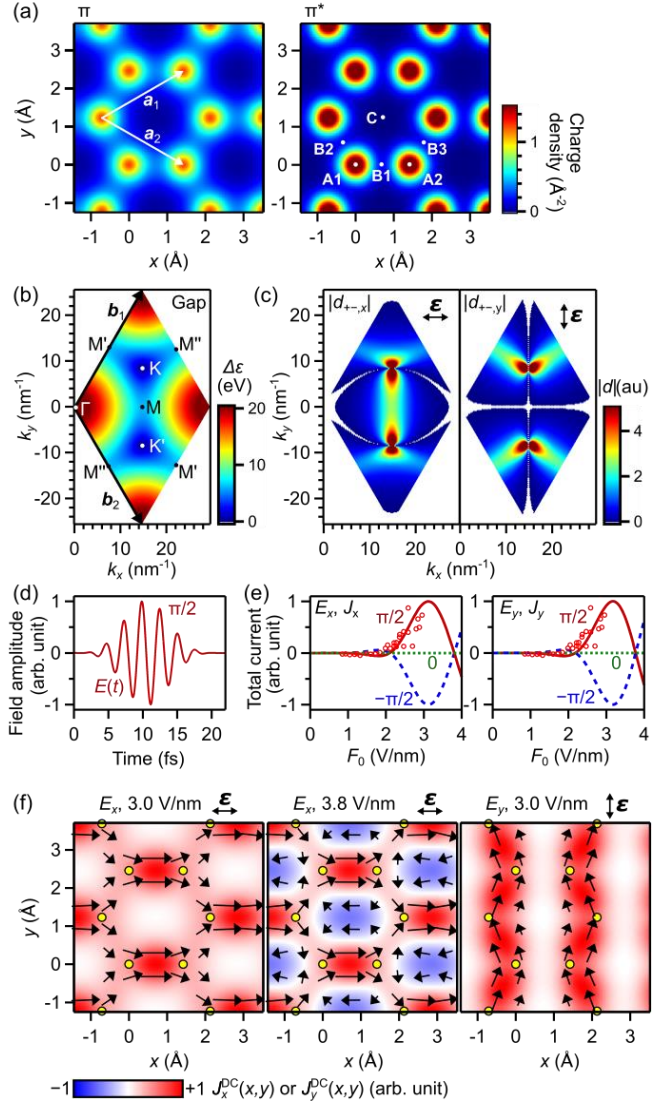

Figure.1. Overview of light-field-driven residual currents in graphene. (a) Charge densities of the π state (left) and the π* state (right) projected onto the *xy* plane. The same color scales are used for the two plots. (b) Band gap energy between the π and the π* states in reciprocal space. (c) Magnitudes of the dipole matrix elements in the Houston basis. (d) The laser electric field waveform at $\varphi_{\text{CEP}} = \pi/2$. (e) Total residual currents with $E_x$ (left) and $E_y$ (right) excitations. The simulated currents (curves) are plotted for the three CEP values as indicated in the panels. For comparison, we plot experimentally observed residual currents[9] by open circles. The experimental results in the two panels are identical because polycrystalline graphene was used[9]. (f) Vectorial real-space images of the light-field-induced residual currents in graphene. CEP is π/2 in all the panels. Black circles filled with yellow represent the positions of nuclei. Left panel: $E_x$ excitation at 3 V/nm peak amplitude. Middle panel: $E_x$ excitation at 3.8 V/nm. Right panel: $E_y$ excitation at 3 V/nm. The color scale and the absolute length of the vectors are normalized in each plot.

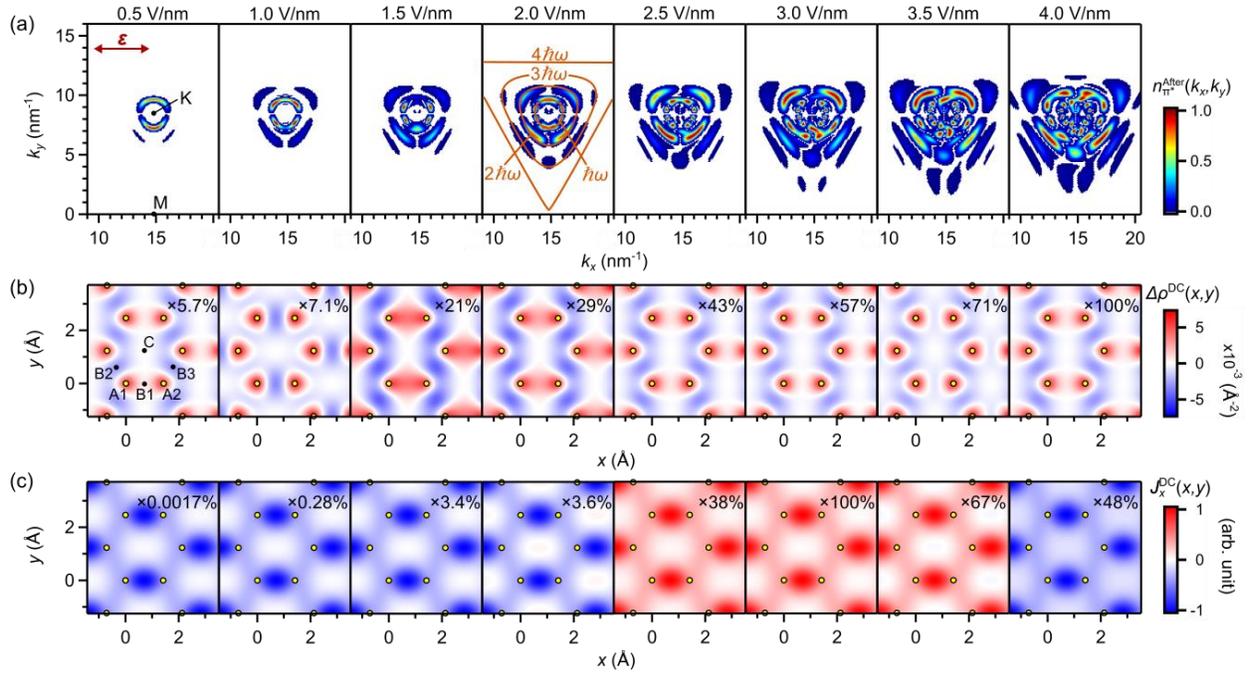

Figure.2. Excited carriers and residual currents driven by $E_x$ field. (a) Reciprocal-space population of the $\pi^*$ state after the laser excitation at variable field strength. A small but clear right-left asymmetry is seen at the field strength of higher than 2.5 V/nm. The up-down asymmetry originates from the band structure, see Fig. 1(b). The orange curves in the plot for 2 V/nm represent wavevectors which give band gaps equal to multiples of photon energy. (b) Changes of the real-space charge densities after the laser excitation. Number in each plot gives the relative color scale. (c) Real-space residual current densities along $x$. The number in each plot shows the relative color scaling factor.

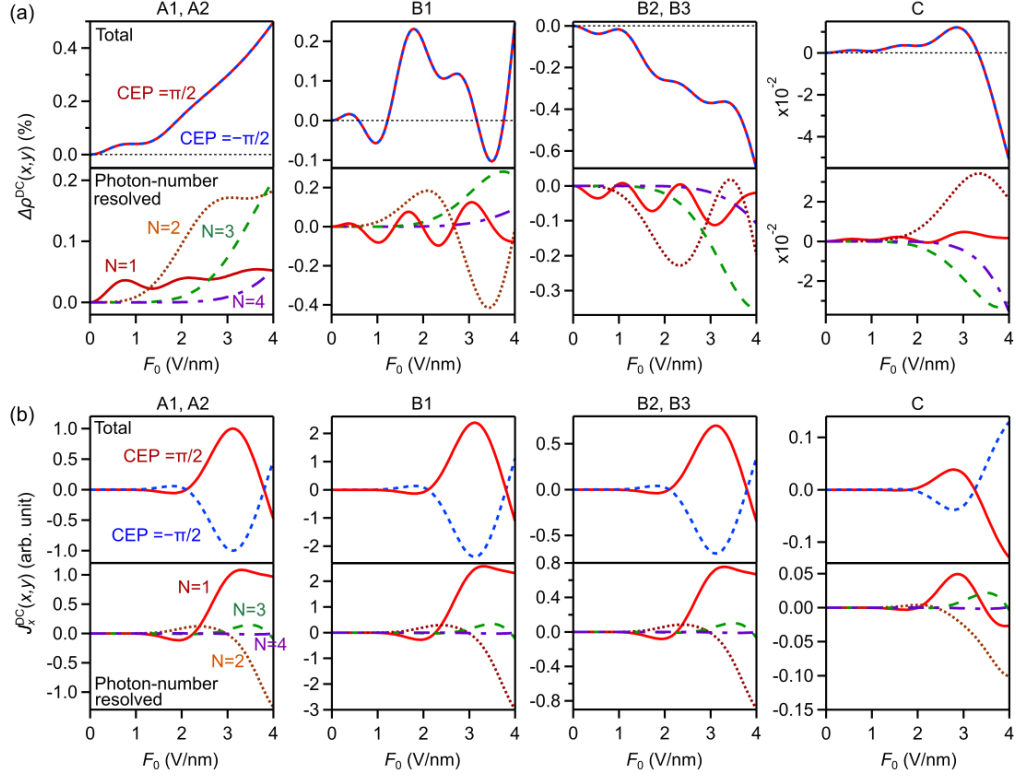

Figure.3. Peak-field amplitude dependence of real-space charge densities and current densities at the six spatial point. (a) Upper panels: change of charge densities normalized by the initial density at each point. Black dotted lines are zero lines. Lower panels: the photon-order decompositions. See main text for details. (b) Upper panels: current densities at each spatial point. The amount of the current is normalized by the largest current density at A1. Lower panels: photon-order decompositions.

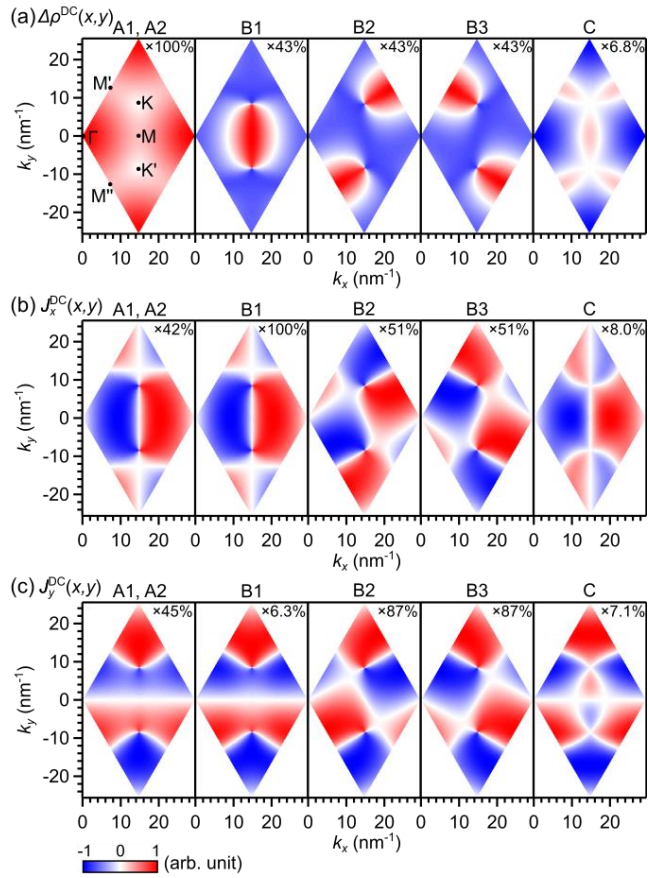

Figure.4. Symmetries in the reciprocal space associated with the real-space charge and current densities. (a) Reciprocal-space symmetries connected with the change of the real-space charge density. The number in each plot gives a color scaling factor. (b) Reciprocal-space symmetries leading to the current densities along $x$ axis. (c) Reciprocal-space symmetries leading to the current densities along $y$ axis. The color scaling facters in (b) and (c) are directly comparable.

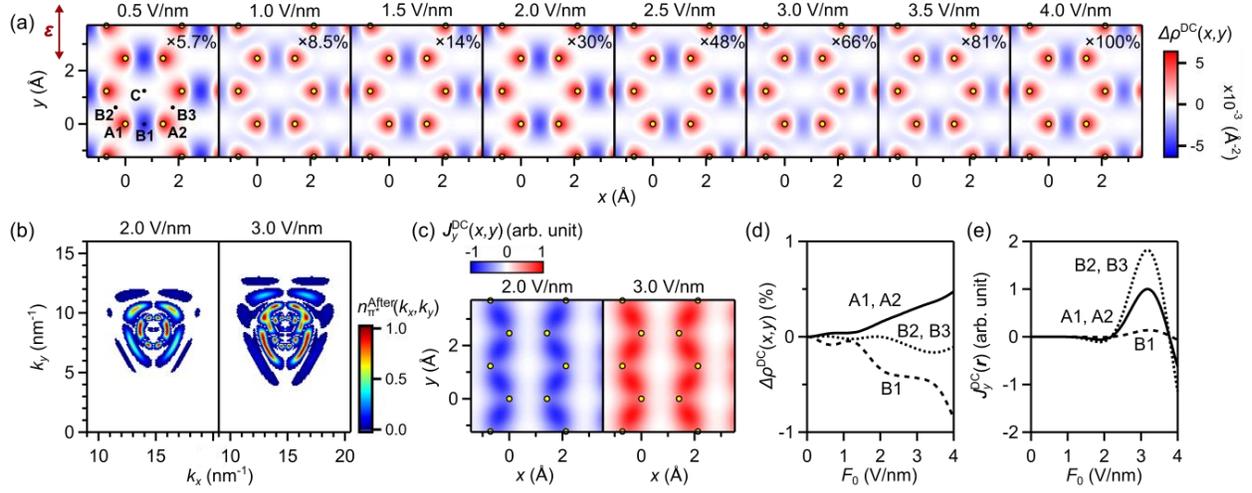

Figure.5. Carrier excitation and residual currents driven by $E_y$ excitation. (a) Changes of the real-space charge densities after the laser excitation. The number in each plot gives the relative color scaling factor. (b) Reciprocal-space carrier populations of the π* state at around the *K* point. (c) Real-space residual current densities. (d) Dependence of charge density changes on the peak field amplitude. (e) Dependence of current densities on the peak field amplitude. The charge and current densities at point C are much smaller than the others and therefore not plotted in (d) and (e).